%% file: main.tex
\documentclass{article}

\usepackage{authblk}
\usepackage[utf8]{inputenc}
\usepackage[margin=1in]{geometry}
\usepackage{multicol}
\usepackage{multirow}
\usepackage{xcolor}
\usepackage{graphicx}
\usepackage{cite}
\usepackage{adjustbox}
\usepackage{float}
\usepackage{amsfonts}
\usepackage{url}
\usepackage{float}
\usepackage{tikzit}
\usepackage{bm}
\input{tree.tikzstyles}
\input{tree.tikzdefs}

\title{Sequence Processing with Quantum Tensor Networks}

\author[1,3]{Carys Harvey} 
\author[2,3]{Richie Yeung}
\author[3]{Konstantinos Meichanetzidis}
\affil[1]{Quantum Engineering Centre for Doctoral Training, University of Bristol, BS8 1UD, UK}
\affil[2]{Department of Computer Science, University of Oxford, OX1 3QD, UK}
\affil[3]{Quantinuum, 17 Beaumont Street, Oxford, OX1 2NA, UK}

\date{\today}

\begin{document}

\maketitle

\begin{abstract}
We introduce complex-valued tensor network models for sequence processing motivated by correspondence to probabilistic graphical models, interpretability and resource compression. Inductive bias is introduced to our models via network architecture, and is motivated by the correlation structure inherent in the data, as well as any relevant compositional structure, resulting in tree-like connectivity. Our models are specifically constructed using parameterised quantum circuits, widely used in quantum machine learning, effectively using Hilbert space as a feature space. Furthermore, they are efficiently trainable due to their tree-like structure. We demonstrate experimental results for the task of binary classification of sequences from real-world datasets relevant to natural language and bioinformatics, characterised by long-range correlations and often equipped with syntactic information. Since our models have a valid operational interpretation as quantum processes, we also demonstrate their implementation on Quantinuum's H2-1 trapped-ion quantum processor, demonstrating the possibility of efficient sequence processing on near-term quantum devices. This work constitutes the first scalable implementation of near-term quantum language processing, providing the tools for large-scale experimentation on the role of tensor structure and syntactic priors. Finally, this work lays the groundwork for generative sequence modelling in a hybrid pipeline where the training may be conducted efficiently in simulation, while sampling from learned probability distributions may be done with polynomial speed-up on quantum devices.


\end{abstract}


\section{Introduction}

In recent years data has become a plentiful resource that has contributed to monumental advances in AI. Application of large language models (LLMs) which are trained on TBs of data, such as GPT-3 \cite{attention, openai}, GPT-4 \cite{openai2023gpt4}, and LAMDA \cite{lamda}, have entered the public sphere and have been met with justifiable awe. However, critiques of this unstructured approach remain prominent and the large amounts of redundancy in these energy-hungry models further encourage us to ask if, despite impressive results, alternative routes are worth exploring. 
 It is believed that naturally occurring learning systems utilise priors, or biases, which provide scaffolding to neural wiring \cite{infant_priors, priors2, bias, bias2, bias3, bias4}.
A motivating question regards the brain's ability to generalise from sparse examples when artificial neural networks typically require such large amounts of data.
\emph{Compositional} generalisation argues that the presence of innate wiring rules predisposes the brain with a model for reasoning and concept creation. Such ideas paved the way for the development of \emph{structured} learning models. The contemporary example is graph neural networks (GNNs) \cite{gnn_review,bronstein2021geometric} which achieve state-or-the-art performance 
with improved sample complexity \cite{gnn_review,bronstein2021geometric, few_shot1,few_shot2, few_shot3, few_shot4}. 

Tensor networks provide low-dimensional representations of high-dimensional data and using them to construct machine learning models has been motivated by the need to reduce redundancy in AI models, the need for \emph{interpretability} \cite{stoudenmire2016supervised, stoudenmire2018learning, novikov2015tensorizing, novikov2016exponential, hallam2017compact, liu2019machine, patel2022quantum}, 
providing rigorous theoretical results such as separations in expressivity \cite{cohen2016expressive, glasser2019expressive}, and their equivalence to probabilistic graphical models\cite{glasser2020probabilistic}.
The focus of this work is natural language processing (NLP) and, more generally, sequence modelling, and tensor networks are a natural playground for modelling symbol sequences.
Firstly, they can natively fuse the successful \emph{distributional} approaches for encoding meaning, or `semantics' \cite{milky1,milky2}, with the \emph{compositional} structures rooted in grammatical and syntactic structures developed in the field of theoretical linguistics \cite{Chomsky57a}.
In the DisCoCat framework \cite{coecke2010mathematical},
the first to introduce such an approach to modelling the semantics of sentences,
word embeddings take the form of tensors
which are composed according to the syntactic structure of the sentence in which they participate,
resulting in a syntax-aware tensor network.
Specifically, the compositional schemes used for constructing DisCoCat models follow parses of typeological grammars, such as pregroup grammar \cite{Lambek58themathematics, lambekword}, Lambek-calculus \cite{LambekLambek2013}, or context-free grammar \cite{Stefano2016}.
Note that in DisCoCat,
the semantics are contained \emph{only} in the word-tensors, and the syntactic structure \emph{only} determines the connectivity between the tensors,
ie the pattern of tensor contractions.
Furthermore, the task of sequence modelling in general reduces to learning a probability distribution over symbols or `tokens'.
There is an exact equivalence between probabilistic graphical models, which are factorisations of probability distributions, and tensor networks, which are factorisations of high-rank tensors into lower-rank tensors \cite{glasser2020probabilistic}.

Tensor networks naturally describe quantum processes \cite{coecke_kissinger_2017}, and so one may consider using quantum Hilbert-space as feature space \cite{SLQuantumFeature2019}, which may lead to superpolynomial advantage for quantum machine learning in principle \cite{RigorousRobust2021,Powerofdata2021}.
Such \emph{quantum tensor networks} \cite{rieser2023tensor} may have valid operational interpretations as quantum computations, for example realised with parameterised quantum circuits (PQCs) \cite{MarcelloReview2019}. This may lead to polynomial speed-up in sampling from the probability distribution on a quantum device, even in the case of efficiently contractable networks \cite{Huggins2019}.
Employing such models for quantum natural language processing (QNLP) can result in polynomial quantum speedup \cite{ZengCoecke2016}, or, beyond computational advantage, one can utilise quantum features such as contextuality \cite{Gao2022,anschuetz2022interpretable}. Specifically, quantum DisCoCat models have been constructed for syntax-aware quantum machine learning tasks \cite{NISQ-QNLP-QPL}, as has been demonstrated in small-scale proof-of-concept experiments \cite{meichanetzidis2020grammaraware,lorenz2021qnlp}.

Regarding network architecture,
one-dimensional models have previously been explored for probabilistic sequence modelling \cite{stokes2019probabilistic, miller2021tensor, glasser2019expressive}. However, from a physical point of view, the presence of long-range correlations (known as Zipf's law) in language and biological data \cite{corr, corr1, corr2, corr3, crit1, crit2} motivates the use of tree-like or hierarchical tensor network architectures \cite{meraLM,lang_renorm,wright2022deterministic}. Key examples of hierarchical tensor networks are the tree tensor network (TTN) \cite{TTNsGM2019} and multi-scale entanglement renormalisation ansatz (MERA) \cite{Vidal2008}, initially introduced to capture power-law correlations in critical quantum many-body systems.
In direct analogue to condensed matter physics we leverage the flexibility in architecture design of tensor networks to restrict the exponentially large feature space according to the correlation structure relevant to a given problem \cite{TNsReviewOrus2019}. That is, we introduce an inductive bias on the space of models according to any known structure in the data and learning task at hand.

Noteworthy explorations of syntax-aware neural-based models include the recursive neural network of Ref. \cite{socher}, defined for the task of sentiment analysis, where the authors motivate the use of syntactic structure from the point of view of interpretability and explainability. Ref. \cite{hermann-blunsom-2013-role} introduces syntax-aware neural-based models where even more linguistic information from a combinatory categorial grammar (CCG) \cite{Steedmanbook} is used, a grammar on which we will base our syntax-aware models in this work, as well. Additionally, improvement in inference and machine translation tasks have been found using syntax-aware neural models in \cite{liu2016syntax, gomez2021syntax}.

In this work, we build models that go \emph{beyond} the DisCoCat framework.
 While we still define models based on compositional structures, and when applicable that structure is given by syntax, we also allow other parts of the model to be trainable  rather than have only the word embeddings be parameterised. We realise a wide range of
quantum-inspired tensor network models for modelling symbol sequences with long-range correlations.
Beyond the description of such rich families of quantum-inspired models in the common language of tensor networks,
one of the main contributions of this work is the introduction of methods for \emph{efficient} implementation of large-scale quantum tree tensor network models. This enables large scale QNLP experiments that importantly do not suffer an exponential cost due to postselection when executed on quantum processors.
Hence this work provides tools for testing the benefits of including syntactic structure in tensor-based models for NLP tasks with real-world data.
Finally, even though in this work we focus on the task of classification, we also sketch how we have laid the groundwork for generative modelling for symbol sequences.

\section{Models}
\label{sec:models}

\begin{figure}[t]
    \centering
    \includegraphics[scale=1.0]{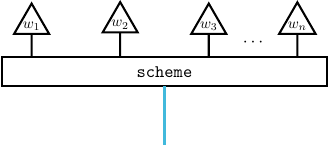} 
    \caption{Process diagram representing a compositional {\tt scheme} for a sentence.
    Black wires carry $\tau$ types and the blue wire carries the $\sigma$ type.
}
\label{fig:scheme}
\end{figure}

We introduce our models via a two-step process.
To instantiate a model,
first, a \emph{compositional scheme} is defined for a sentence.
In order to define a compositional scheme,
we make use of the graphical language of process theories.
The processes are depicted by \emph{boxes}, with input and output \emph{wires}.
The compositional schemes we use throughout this work are such that wires carry one of two types, the `internal' type $\tau$ and the `sentence' type $\sigma$,
as shown in Figure \ref{fig:scheme}.
Freely composing the boxes, while respecting the types, allows for any process diagram to be generated, representing a {\tt scheme}, to be defined.
Given a vocabulary comprising a finite set of words, or more generally tokens, $V=\{w_i\}_i$, we consider compositional schemes for sentences, or more generally sequences, $S$, of finite length over this vocabulary.

\begin{figure}[t]
    \centering
    \includegraphics[width=\textwidth]{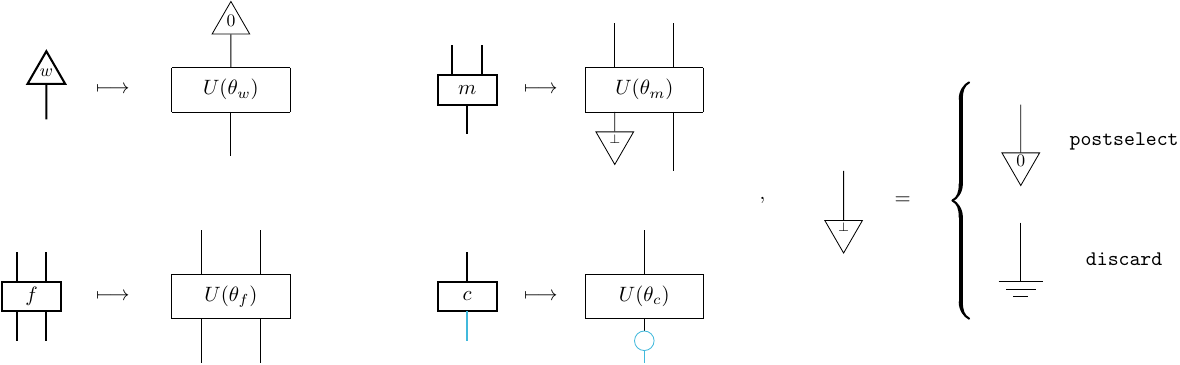} 
    \caption{Definition of the semantic functor $F$ which assigns parameterised quantum circuits $U(\theta_b)$ to boxes of type $b$.
    Thick black wires carry $\tau$ types and via $F$ they are mapped to thin black wires carrying $q$-many qubits.
    The thick blue wire carries the $\sigma$ type and is mapped to a thin blue wire carrying $b$-many bits.
    For the $\bot$-effect, we can choose either to {\tt postselect} on the all-zeros state or to {\tt discard}.
}
    \label{fig:functor}
\end{figure}

Then, semantics is given via a \emph{semantic functor}, $F$, i.e. a structure-preserving map.
The semantic functor $F$ gives Hilbert space semantics to our compositional schemes, realising quantum tensor network models. This is done by assigning a number of qubits $q$ to $\tau$ and $q'$-many bits to $\sigma$, and a parameterised quantum circuit (PQC) of a suitable size to each box, where the set of control parameters depends on the box:
\begin{itemize}
    \item The $w$-box, which specifically prepares a \emph{word-state}, typed $\tau^{\otimes 0} \to \tau^{\otimes 1}$, is mapped to a parameterised quantum state prepared by applying the circuit $U(\theta_w)$ on the fixed input state $|0\rangle^{\otimes q}$. Note that to every word corresponds a parameter set.
    \item The \emph{filter} $f$-box, with type $\tau^{\otimes 2} \to \tau^{\otimes 2}$, is mapped to $U(\theta_f)$ with $2q$ input and $2q$ output qubits.
    \item The \emph{merge} $m$-box, typed $\tau^{\otimes 2} \to \tau^{\otimes 1}$, is mapped to a $U(\theta_m)$ with $2q$ input qubits and $q$ output qubits, where the other $q$ qubits have been either discarded or postselected by the $\bot$-effect.
    \item The \emph{classifier} $c$, with type $\tau^{\otimes 1} \to \sigma^{\otimes 1}$, is mapped to a $U(\theta_c)$ which accepts a $q$-qubit state as input and outputs a $q'$-qubit state (by applying the $\bot$ effect), which is then measured in the $Z$ basis to get a vector in $[0, 1]^{2^{q'}}$ which represents the probability of each outcome determined by the Born rule.
\end{itemize}
The action of $F$ on the scheme generators is shown in Figure \ref{fig:functor}.
In general, a more general class of quantum processes can be considered by introducing ancillae qubits to the parameterised quantum circuits.

For the $\bot$-effect, we consider two options.
In the first case, $\bot={\tt postselect}$, it represents postselection, ie conditioning on the outcome of a nondeterministic measurement, by convention and without loss of generality to the all-zeros state.
In the second case, $\bot={\tt discard}$, it represents discarding of that dimension, or a partial trace (marginalisation).
The tensor network topology of the models we consider, inherited by the structure of the compositional schemes we define, allows for \emph{efficient} tensor contraction strategies, thus making them classically simulable.
This holds for both choices regarding the $\bot$-effect.
We can therefore consider our models as `quantum inspired'.
At the same time, both versions have a valid operational interpretation as quantum circuits, so they could also be evaluated on quantum processors.
However, while discarding is a \emph{free} operation in quantum theory, postselection is not, as it incurs an \emph{exponential} sampling overhead in number of qubits being postselected.

The number of qubits per wire $q$,
the specific form, or ansatz, of each PQC realising a unitary $U$, and the choice for the $\bot$-effect are \emph{hyperparameters} of our models.
The ans\"{a}tze we use are composed of a number of repeating layers of parameterised quantum gates.
For simplicity, our functor is defined such that
all PQCs involved are defined in terms of the same ansatz $U(\theta_b)$ and have the same number of layers $D$; the type of box $b$ affects the control parameters $\theta_b$ as well as the size of the circuit.
In the following, we taxonomise our tensor network models first by the compositional {\tt scheme} defining their architecture,
and then by the choice of semantic functor $F(\theta)$ and its hyperparameters, where $\theta = \cup_b \theta_b$ is the set of all control parameters (in the following we will suppress notation for $F$ as it should be inferable by context).
Rules and constraints we impose on the parameters that control PQCs define a model \emph{species}.

\subsection{Path}

\begin{figure}[t]
    \centering
\includegraphics[scale=0.8,align=c,draft=false]{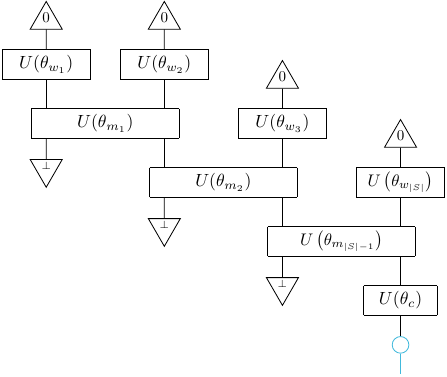}
\includegraphics[scale=0.8,align=c,draft=false]{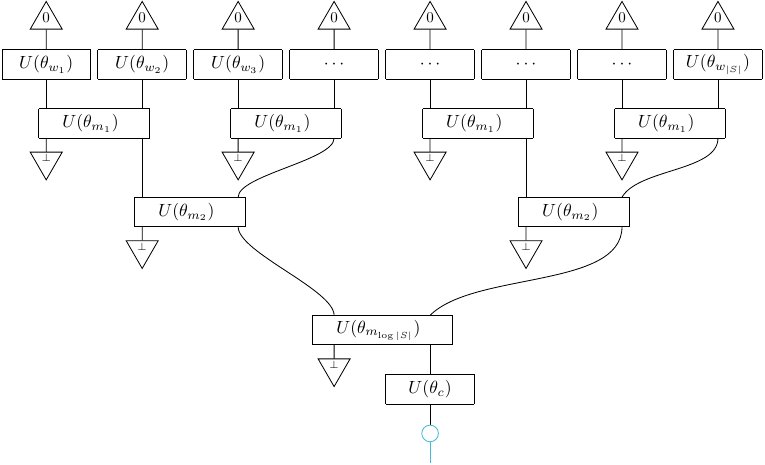}
\includegraphics[scale=0.8,align=c,draft=false]{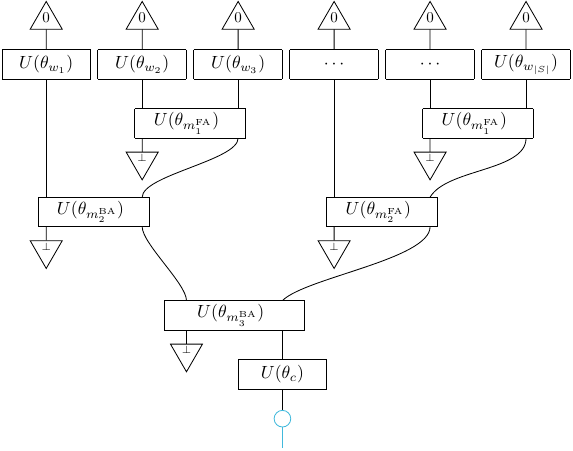}

    \caption{The scalable models: Path tensor network $F(\theta)[{\tt path}]=\mathrm{PTN}$ (top-left), tree tensor network $F(\theta)[{\tt tree}]=\mathrm{TTN}$ (top-right), and syntactic tensor network $F(\theta)[{\tt syntax}]=\mathrm{STN}$ for a given syntactic structure (bottom).}
    \label{fig:scalable_models}
\end{figure}

The first compositional scheme we study follows a sequential architecture which follows the \emph{reading order} of the words in the sentence.
Applying the semantic functor $F$
defined in Figure \ref{fig:functor}
to this scheme,
we obtain the path tensor network model $F(\theta)[{\tt path}] =\mathrm{PTN}$,
as shown in Figure \ref{fig:scalable_models}.

We denote the length of the input sequence $S$ by $|S|$. If we allow a `hierarchical' dependence of the parameter sets $\theta_{m_i}$
on their position $i\in \{1,2,\dots,|S|-1\}$, then we realise the hPTN models.
Then imposing that all $m$-circuits share one parameter set $\theta_m=\theta_{m_i}, \forall i$, which makes the model recurrent,
we realise the `uniform' models uPTN.
The uPTN
can be interpreted as a vanilla version of a recurrent quantum model \cite{bausch2020recurrent}, or a matrix product state (MPS) model where all the `physical dimensions' are either discarded or postselected, save the last one which encodes the meaning of the sentence.

This compositional scheme does not take syntax nor long-range correlations into account and can thus serve as a baseline model against which syntax-aware models can be compared, but still satisfies the property that it is constructed by local application of a composition rule.
Also, as it is effectively an MPS, it is not expected to be able to capture long-range correlations for a fixed bond dimension,
constituting it a useful baseline against other models we construct below.

\subsection{Tree}

The simplest model that is able to capture long-range correlations uses a compositional scheme with the topology of a \emph{balanced binary tree}.
We call this scheme {\tt tree}.
At every branching, an $m$-box combines two input wires into one. The number of branchings, and so the number of $m$-boxes is $|S|-1 = \sum_{i=0}^{{\log(|S|)-1}}2^i $, where the log is in base 2 throughout. This scheme has depth $\log(|S|)$, since at the $i$-th layer there are half as many parallel $m$-boxes than that in layer $i+1$.
Figure \ref{fig:scalable_models} shows the result of applying the semantic functor to {\tt tree} in order to obtain a tree tensor network (TTN) model.

We explore the `hierarchical' species, hTTN, where circuits that belong to the $i$-th layer share a parameter set $\theta_{m_i}$.
Furthermore, we define the `uniform' species, uTTN, where all $m$-circuits share one parameter set $\theta_m=\theta_{m_i},\forall i$.
The general form of quantum TTN is what is defined in Ref. \cite{Huggins2019}, where the authors allowed the $m$-circuits to have different parameter sets and used this model to classify images of handwritten digits (MNIST).

\subsection{Syntactic}

We now present the first model that is syntax-aware, which we call the {\tt syntax}.
This compositional scheme is a binary tree
whose branching structure is
given by the sentence's \emph{syntactic structure}
as output by a CCG parser.
Structure-wise, it can be viewed as an intermediate scheme between {\tt path} and {\tt tree}, in the sense that the CCG parser may output binary trees that include {\tt path} (right-branching or maximally-imbalanced binary tree) and {\tt tree} (ie balanced binary tree), with {\tt path} the limit case of maximal depth and minimal width and {\tt tree} the limit case of minimal depth and maximal width.
The number of $m$-boxes in this scheme is also $|S|-1$,
but the depth depends on the syntactic structure.
In Figure \ref{fig:scalable_models} we an example of an instantiation of a syntactic tensor network (STN).

More specifically,
we define three species of STNs.
The $m_i^r$-circuits depend on the layer $i\in\{1,2,\dots,|S|-1\}$, ie their distance from the leaves, as well as the CCG rule $r\in \mathrm{R}$ as annotated by the parser. 
Similarly to the case of models based on the {\tt tree} scheme, we can also, in this case, define a hierarchical species
(hSTN),
such that the $m$-circuits that are at depth $i$, where depth is defined by their distance from the leaves, share a parameter set $\theta_{m_i}=\theta_{m_i^r}, \forall r$.
Also, since we have syntactic information,
we can define a `rule-based' species (rSTN),
where the $m$-circuits annotated with the same CCG rule $r$ share the same parameter set $\theta_{m^r} = \theta_{m_i^r}, \forall i$.
Finally, we define the
uniform species (uSTN),
where all $m$-circuits share a single parameter set $\theta_{m}=\theta_{m_i^r}, \forall i,r$.

\subsection{Convolutional}

\begin{figure}[t]
\centering
$$
\includegraphics[scale=0.8,align=c,draft=false]{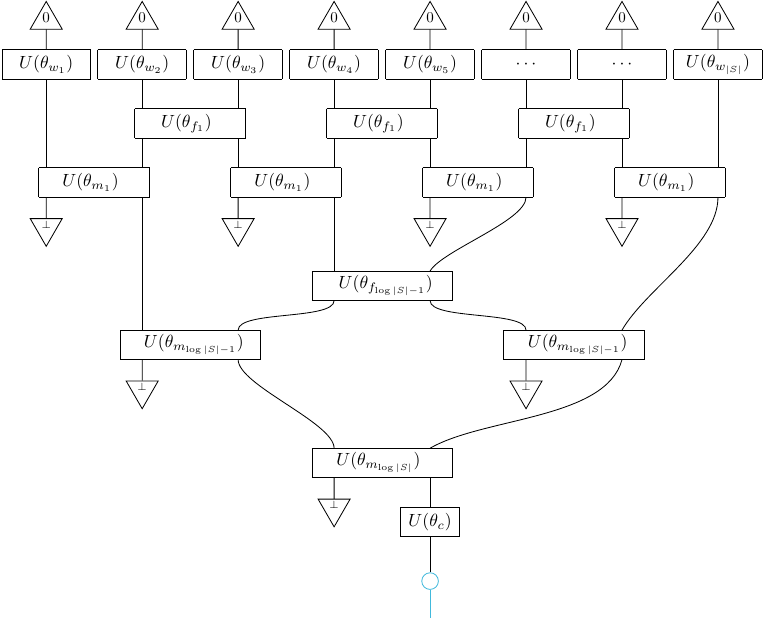}
$$
    \caption{Convolutional tensor network $F(\theta)[{\tt conv}]=\mathrm{CTN}$.}
    \label{fig:ctn}
\end{figure}

We now introduce a compositional scheme that can be viewed as an enhancement of the {\tt tree}, which we call {\tt conv}.
It is constructed by adding layers of $f$-boxes to the {\tt tree} scheme, such that they act before the $m$-boxes on neighbouring wires that are not input to the same $m$-box.
The number of $m$-boxes is $|S|-1 $ and the depth is $\log(|S|)$ as is the case for {\tt tree}.
The number of $f$-boxes is $|S|-(\log(|S|)+1) = \sum_{i=0}^{\log(|S|)-1}(2^i-1)$,
since at every layer there is one less $f$-box than there are $m$-boxes.
The application of the semantic functor to this scheme in order to obtain \emph{convolutional} tensor network $F[{\tt conv}] = \mathrm{CTN}$ is shown in Figure \ref{fig:ctn}.
At the $i$-th layer, the layer of $f$-circuits filter out unnecessary entanglement and then the layer of $m$-circuits reduces the number of qubit wires, effectively coarse-graining the sentence and retaining only the relevant information for the task.

We define the `hierarchical' models hCTN
by having their parameter sets,
$\theta_{f_i}, \theta_{m_i}$,
be shared if they belong to the same layer.
Finally, by making them be shared throughout the model, $\theta_f = \theta_{f_i}, \theta_m = \theta_{m_i}, \forall i$, we realise the self-similar `uniform' uCTN.
The CTN with $\bot = {\tt discard}$ has been introduced in Refs. \cite{Marcello2018,cong} and implemented on quantum computers in Ref. \cite{herrmann2021realizing}.
In Ref. \cite{Marcello2018} the authors put no constraints on the parameter sets $\theta_f, \theta_m$ and the model was used to classify MNIST,
and in Ref. \cite{cong} the authors introduce a special case of hCTN, where measurement-and-feedforward is used to keep the state pure.
Finally, we define a `sliding' variant which allows us to apply the convolutional model to large sequences efficiently. We call this the CTNS which consists of a sliding window of fixed size $w$ that scans along the sequence. At each position in the sequence, $i$, a CTN is applied to the sub-sequence $[w_i,...,w_{i+w}]$.
The outputs obtained from all sub-sequences are then aggregated by a function of our choosing, and here we choose the average.


\subsection{Syntactic Convolutional}

\begin{figure}[t]
    \centering
$$
\includegraphics[scale=0.8,align=c,draft=false]{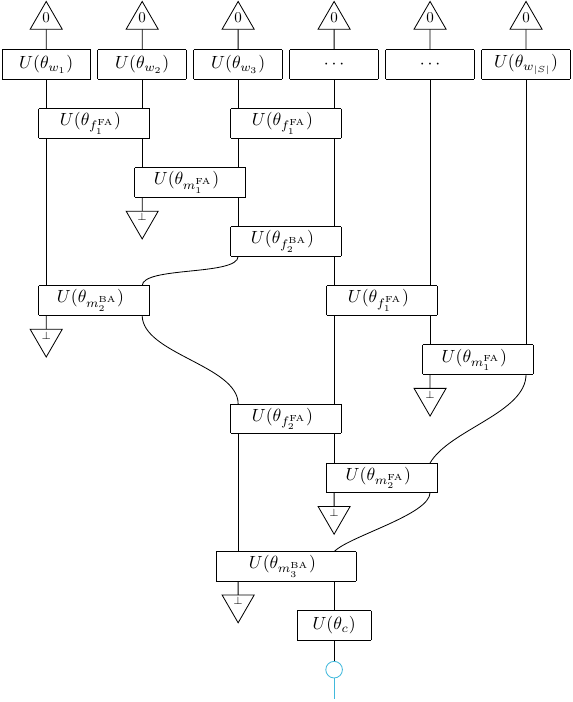}
$$
    \caption{Syntactic convolutional tensor network $F(\theta)[{\tt syntaxconv}]=\mathrm{SCTN}$, for a given syntactic structure.}
    \label{fig:sctn}
\end{figure}

Finally, we introduce a compositional scheme
that combines the features of {\tt syntax} and {\tt conv}, i.e. it is a syntax-aware coarse-grainer of a sentence, which we call {\tt syntaxconv}.
It is constructed by endowing {\tt syntax} with $f$-boxes, following the convention that before every $m$-box 
we place $f$-boxes.
The rule for placing an $f$-box
is that it does not act on two wires that
are inputs to the same $m$-box.
In general, $f$-boxes do not necessarily commute,
so we follow the convention that if two $f$-boxes act on the same wire, then the one that involves words that appear earliest in the sentence is applied first.
The application of the semantic functor which returns the syntactic convolutional tensor network, $F[{\tt syntaxconv}] = \mathrm{SCTN}$, is shown in Figure \ref{fig:sctn}.


The $m$-circuits and $f$-circuits are controlled by parameter sets that depend on the CCG rule annotating the tree at each branching,
as well as the depth (distance from the leaves), $\theta_{m_i^r}, \theta_{f_i^r}$.
We define the `hierarchical' models hSCTN, where circuits of the same depth share the same parameter sets $\theta_{m_i} = \theta_{m_i^r}$ and $\theta_{f_i} = \theta_{f_i^r}$, $\forall r$.
We also define the `rule-based' models 
rSCTN, in which case the parameter sets only depend on the CCG rule
$\theta_{m^r} = \theta_{m_i^r}$ and $\theta_{f^r} = \theta_{f_i^r}$, $\forall i$.
Finally, the uniform models uSCTN have two parameter sets shared across all $m$-circuits and $f$-circuits,
$\theta_{m} = \theta_{m_i^r}$ and $\theta_{f} = \theta_{f_i^r}$, $\forall i,r$.

\section{Experimental results for binary classification}
\label{sec:results-classification}

The models as we have defined them in Section \ref{sec:models} follow an `encoder' architecture, in that they take words from a sentence as input, and return an output.
Thus, they are designed to naturally act as \emph{classifiers}.
We focus on tasks of binary classification,
ie only a single qubit ($q'=1$) is measured from the output quantum state of the $U_c$ circuit, resulting in the probabilities over two measurement outcomes, $p_0$ and $p_1$, as given by the Born rule.
Each outcome is interpreted as the label of a class.
Multiclass classification can be straightforwardly defined by setting $q' = \lceil \log L \rceil$ for a number of class labels $L$.
Furthermore, since all the model species we have introduced have a \emph{tree-like} architecture, they can be evaluated \emph{efficiently}; we show the upper bounds to their contraction complexities in Table \ref{tbl:complexities}.
Furthermore, tree-like models are immune by construction to the phenomenon of barren plateaus during training \cite{noBPs,Enrique2023, zhao2021analyzing}.

\begin{table}[H]
\centering
\begin{adjustbox}{width=0.75\textwidth}
\small
\begin{tabular}{ |c|c|c|c|c|c| } 
\hline
$\bot$ & PTN & TTN & STN & CTN & SCTN  \\
\hline
{\tt discard} &  $\mathcal{O}(|S|\chi^4)$ & $\mathcal{O}(|S|(\chi^2)^3)$ & $\mathcal{O}(|S|(\chi^2)^3)$ & $\mathcal{O}((\chi^2)^{2\log{(|S|)}})$ & $\mathcal{O}((\chi^2)^{2\log{(|S|)}})$    \\ 
{\tt postselect} & $\mathcal{O}(|S|\chi^3)$ & $\mathcal{O}(|S|\chi^3)$ & $\mathcal{O}(|S|\chi^3)$ & $\mathcal{O}(\chi^{2\log{(|S|)}})$ & $\mathcal{O}(\chi^{2\log{(|S|)}})$  \\ 
\hline
\end{tabular}
\end{adjustbox}
\caption{Contraction complexity upper bounds for quantum and classical models as a function of the wire dimension $\chi=2^q$ and the length of the sequence $|S|$.}
\label{tbl:complexities}
\end{table}

The models are defined using the \texttt{DisCoPy} library for monoidal categories \cite{de2020discopy, toumi2022discopy}.
Where applicable, the syntactic structures were obtained
using the CCG parser {\tt Bobcat} \cite{clark2021old}, which is available in the {\tt lambeq} package \cite{kartsaklis2021lambeq}.
We simulate the models using the \texttt{tensornetwork} library \cite{roberts2019tensornetwork}.
Training is done using the library JAX \cite{JAX2018github}; by defining the forward pass as a pure JAX function, we are able to exploit Just-In-Time (JIT) compilation to batch compute the outputs of multiple sentences, even if the tensor networks were produced from schemes with different syntactic structures. While we experimented with different families of quantum ans\"{a}tze, the results we present use the expressive ansatz 14 from \cite{sim2019expressibility}, as it had the best test performance.
The parameters of a model are optimised using AdamW \cite{AdamW} such that the labels of the train set $l_i\in\{0,1\}$ are predicted correctly according to the binary cross-entropy loss, $H = - \frac{1}{|\mathrm{train ~set}|} \sum_{i=1}^{|\mathrm{train ~set}|} l_i \log_2({p_1}_i) + (1-l_i) \log_2(1-{p_1}_i)$.
If one were to train on a quantum computer, one would use the parameter-shift rule for estimating analytic gradients \cite{ParameterShift2019}, or use gradient-free optimisers such as SPSA \cite{SPSA119632}, which requires much less overhead and has been shown to perform well when training PQCs on near-term hardware \cite{PerformanceComparisonOptimisers2023}.
A validation set is used for model selection with early stopping. The model hyperparameters are the embedding qubit number and ansatz depth, $(q,D)$, and learning rate. The optimiser is given a fixed random seed for all models. Finally, an unseen test set is used to evaluate the model's generalisation performance as measured by the prediction accuracy.
 All test accuracy results are taken at best validation accuracy over the stated hyperpameters and learning rate with fixed seed rather than an averaged behaviour.

We work with three datasets comprising labelled sequences: two NLP datasets, one for detecting whether short news titles are clickbaits \cite{clickbait} and one for sentiment analysis on Rotten Tomatoes movie reviews\cite{RT}, as well as a third dataset relevant to bioinformatics, for classification of the binding affinity of DNA sequences \cite{zou2018primer}.
The NLP data is lemmatised, which occurs post-parsing for the syntactic models.
While PTN, TTN, STN are all efficiently batchable and scalable,
CTN and SCTN remain harder to scale in simulation, as discussed in Appendix B. Tables \ref{tbl:scalable-clickbait},\ref{tbl:scalable-RT} show results only for the scalable models on the full NLP datasets.
In Table \ref{tbl:allmodels-reducedclickbait}, we present results from all models where we have restricted $q=1$. In order to obtain results from SCTN, we define the `reduced' Clickbait dataset by implementing batching by grouping the sentences with the same syntactic structure and utilising vectorisation.
In Table \ref{tbl:scalable-DNA}, we present results for the DNA binding dataset from the scalable models, excluding the syntax-aware STN since we do not make use of a parser for the DNA strings.
In Table \ref{tbl:allmodels-DNA}, we show results for DNA binding from all non-syntactic models restricted to $q=1$ in order to accommodate the relatively more costly CTN model.

As expected it is not possible to show any significant advantage for including syntax for sentiment tasks. It is however noteworthy that even for a sentiment task the CTNS outperforms all models on the Rotten Tomatoes dataset. Additionally, for the genetic data set with models restricted to $q=1$ the CTNS performs well but the complete CTN significantly outperforms all other models. This suggests the importance of an architecture's bias towards the correct correlations in achieving resource compression. For the uCTN achieving 89.0$\%$ test, the number of trainable parameters is 3/token and 11 in the remaining architecture. For the hCTN achieving 94.0$\%$ test, the number of trainable parameters is 3/token and 8/layer in the remaining architecture (with 4 token and 6 layers in total) and again 3 for the final classification box. Note that for all models presented the number of trainable parameters are very small (see Table 7 in Appendix A for exact values) compared to standard NLP models.

We additionally ran the uCTNS model with $\bot={\tt postselect}$, with window size 4 and ansatz depth 1 on a 50,000 review IMDb dataset, achieving a test set accuracy of 88$\%$. We did not do a full hyperparmeter grid search for this data but this is comparable to recent baseline results \cite{imdb} again with only 3 trainable parameters per word and 11 in the remaining architecture.

 \begin{table}[H] 
\centering
\begin{adjustbox}{width=0.8\textwidth}
\small
\begin{tabular}{ |c|c|c|c|c|c|c|c| } 
\hline
$\bot$ & uPTN & hPTN & uTTN & hTTN & uSTN & hSTN & rSTN \\
\hline
{\tt discard} & 98.2 (3, 1) & 98.7 (3, 1) & 98.6 (1, 2) & 98.0 (1, 2) & \textbf{98.9} (1, 2) & 98.7 (1, 2) & 98.6 (1, 2) \\ 
{\tt postselect} & 99.2 (1, 2) & 99.4 (1, 2) & 98.7 (1, 2) & 99.3 (1, 1) & \textbf{99.5} (1, 1) & 99.2 (1, 2) & 99.4 (1, 1) \\ 
\hline
\end{tabular}
\end{adjustbox}
\caption{Test Accuracy on the Clickbait dataset for the best choice of hyperparameters $(q,D)$ where $q=1,2,3$ and $D={1,2}$. Train/val/test : 25,312/3,165/3,165}
\label{tbl:scalable-clickbait}
\end{table}

 \begin{table}[H]
\centering
\begin{adjustbox}{width=1.0\textwidth}
\small
\begin{tabular}{ |c|c|c|c|c|c|c|c|c|c| } 
\hline
$\bot$ & uPTN & hPTN & uTTN & hTTN & uSTN & hSTN & rSTN & uCTNS & hCTNS \\
\hline
{\tt discard} & 72.6 (1, 1) & 71.1 (1, 1) & 72.7 (1, 2) & 66.5 (1, 2) & 69.8 (1, 2) & 67.8 (1, 2) &  70.7 (1, 2) & \textbf{76.76} (4, 2) & 75.63 (4, 2) \\ 
{\tt postselect} & \ 74.3 (1, 1) & 68.0 (2, 1) & 70.7 (1,1) & 67.1 (2,2) & 72.8 (1, 2) & 67.2 (1, 2)  & 70.5 (1, 2) & \textbf{75.16} (4, 1) & 74.51 (4, 2) \\ 
\hline
\end{tabular}
\end{adjustbox}
\caption{Test Accuracy on the Rotten Tomatoes dataset for best choice of hyperparameters. For the CTNS model $q=1$ only and hyperparameters are $(w,D)$, where $w={4,8}$ and $D={1,2}$. For all other models hyperparameters are $(q,D)$, where $q=1,2,3$ and $D={1,2}$. Train/val/test: 8,362/1,046/1,046}
\label{tbl:scalable-RT}
\end{table}


\begin{table}[H]
\centering
\begin{adjustbox}{width=0.525\textwidth}
\small
\begin{tabular}{ |c|c|c|c|c|c| } 
\hline
$\bot$ & uCTN & hCTN & uSCTN & hSCTN & rSCTN \\
\hline
{\tt discard} & \textbf{96.0 (2)} & 95.0 (1) & 92.7 (1) & 86.0 (1) & 93.1 (1) \\ 
{\tt postselect} & \textbf{97.0 (1)} & 94.9 (2) & 96.3 (2) & 82.8 (1) & 90.2 (2) \\ 
\hline
\end{tabular}
\end{adjustbox}
\caption{Test Accuracy on the \textit{reduced} Clickbait dataset for $q=1$ only,  for the best choice of hyperparameter $(D)$ for $D=1,2$. Train/val/test: 2,033/763/763}
 \label{tbl:allmodels-reducedclickbait}
\end{table}

\begin{table}[H] 
\centering
\begin{adjustbox}{width=0.535\textwidth}
\small
\begin{tabular}{ |c|c|c|c|c| } 
\hline
 $\bot$ & uPTN & hPTN & uTTN & hTTN\\
\hline
{\tt discard} & \textbf{95.0 (3, 2) }& 85.5 (3, 2) & 74.0 (1,2) & 87.5 (2,2) \\ 
{\tt postselect} & 72.5 (1, 1)  & 71.0 (2, 2) & 81.5 (2, 2) & \textbf{92.0 (2,1)} \\ 
\hline
\end{tabular}
\end{adjustbox}
\caption{Test Accuracy on the DNA binding dataset for the best choice of hyperparameters $(q,D)$, where $q=1,2,3$ and $D={1,2}$. Train/val/test : 1,600/200/200}
\label{tbl:scalable-DNA}
\end{table}

\begin{table}[H]
\centering
\begin{adjustbox}{width=0.85\textwidth}
\small
\begin{tabular}{ |c|c|c|c|c|c|c|c|c|c| } 
\hline
$\bot$ & uPTN & hPTN & uTTN & hTTN & uCTN & hCTN & uCTNS & hCTNS\\
\hline
{\tt discard} & 71.0 (2)  & 70.0 (1) & 74.0 (2) & 72.5 (2) & 72.5 (1) & 80.0 (2) & 76.0 (4, 2) & \textbf{81.0} (8, 2) \\ 
{\tt postselect} & 72.5 (1) & 70.0 (1) & 74.0 (2) & 80.5 (1) & 89.0 (1) & \textbf{94.0} (1) & 79.0 (8, 2) & 80.0 (4, 2) \\ 
\hline
\end{tabular}
\end{adjustbox}
\caption{Test Accuracy on the DNA binding dataset for $q=1$ only, for the best choice of hyperparameters $(w,D)$ for the CTNS models where $w={4,8}$ and $D={1,2}$ and $(D)$ with $D={1,2}$ for all other models. Train/val/test : 1,600/200/200}
\label{tbl:allmodels-DNA}
\end{table}



\subsection{Execution on a trapped-ion quantum processor}

\begin{figure}[t]
    \centering
    \includegraphics[width=1\textwidth]{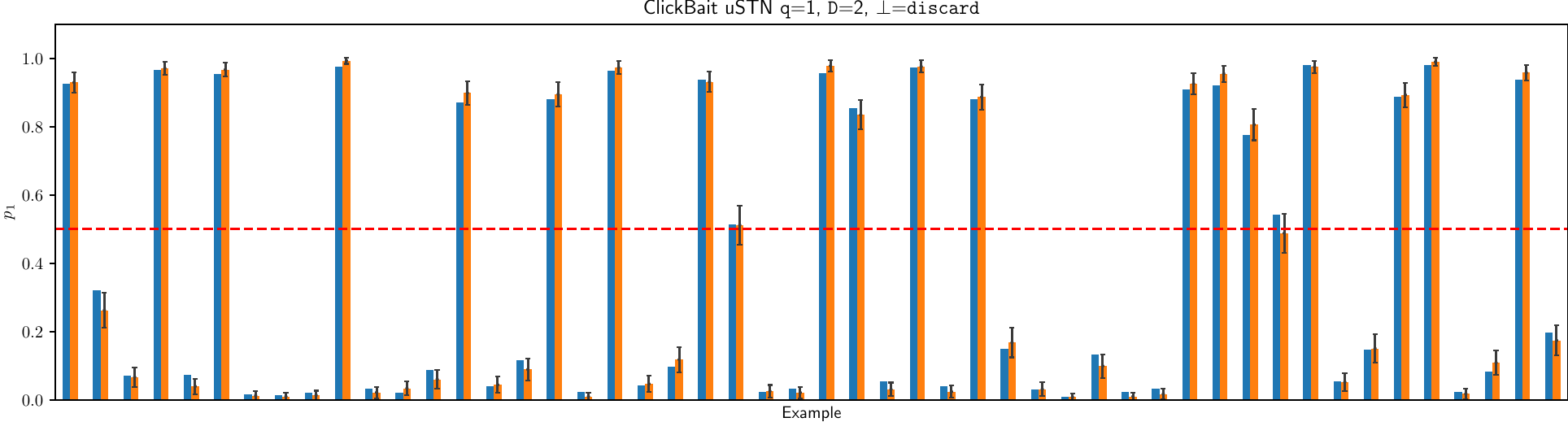}
    \vspace{1em}
    \includegraphics[width=1\textwidth]{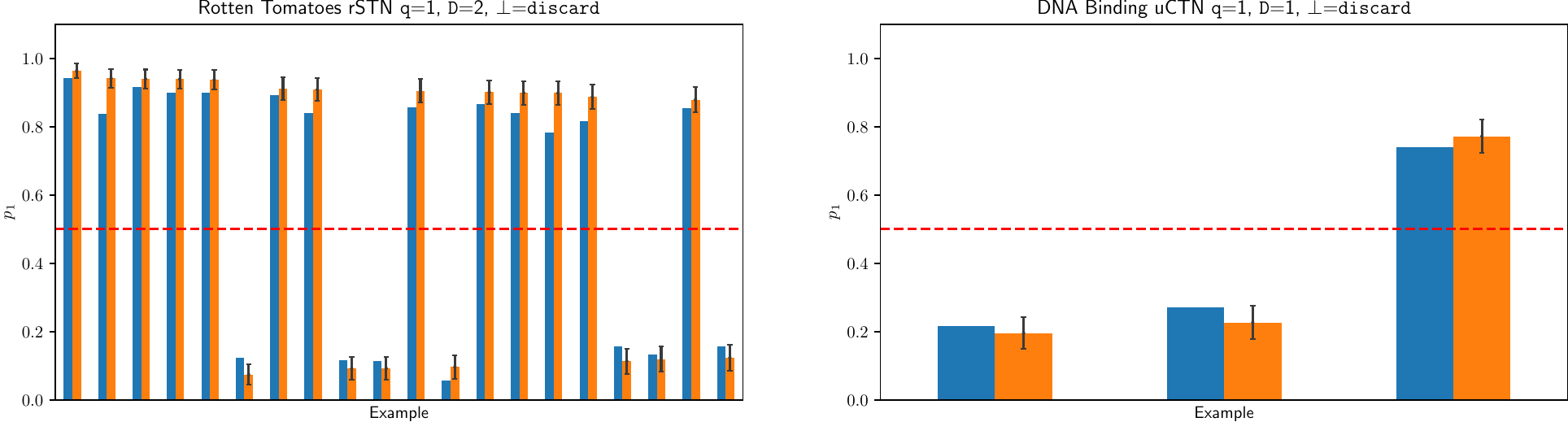}
    \caption{Test-time results for representative examples from the datasets used in this work for some choice of models. The sampled average $\tilde{M} = \frac{1}{300} \sum_{i=1}^{300} M_i$ (blue), obtained by execution on the H2-1 quantum processor for 300 shots, and the expected value $\mathbb{E}(M)$ (orange), obtained by exact classical noiseless simulation, of the measurement outcome $M$, which is a Bernoulli random variable. Specifically, $p_1 = \mathbb{P}(M = 1) = \mathbb{E}(M)$ represents the probability of measuring $1$, and the red line represents the decision threshold ($0.5$). Error bars show the range of 2 standard deviations of $M$ (95\% confidence interval) away from the expected value $\mathbb{E}(M)$, considering only shot noise.}
    \label{fig:H2-results}
\end{figure}

Finally, we execute a representative selection of test set examples from the Clickbait, Rotten Tomatoes and DNA binding datasets, using the uSTN, rSTN and uCTN models with $\bot = {\tt discard}$ respectively, on Quantinuum's 32-qubit H2-1 quantum processor with reported quantum volume of $2^{16} $\cite{H2paper}.
The results are in good agreement with the simulations, as shown in Figure \ref{fig:H2-results}.
For the shorter circuits relevant to classifying the Clickbait dataset, we find that the estimated probability $p_1$ that determines the classification label agrees with its exact classical noiseless simulation, considering the variance arising from shot noise.
From the 50 Clickbait circuits executed on H2-1, we find that only 1 disagrees with its classical simulation regarding the classification label, due to both shot noise and the fact that for that circuit the exact result for $p_1$ is very close to the classification threshold.
For the selected larger circuits corresponding to sequences from the Rotten Tomatoes and DNA binding datasets,
we find that noise effects beyond shot noise, such as gate noise, have affected the estimated value of $p_1$. However, after considering the size of these circuits, it is notable that the classification label is not affected, since noise has not altered $p_1$ enough for it to cross the classification threshold.
Before the quantum circuits corresponding to these models can be run on a specific backend,
they need to be compiled, for which we used Quantinuum's TKET compiler \cite{tket2020}.
This entails translating the circuit's gates into the native operations available to the device.
The quality of the experimental results in relation to exact classical simulation is particularly impressive for the DNA binding sequences comprising 64 nucleotides each.
In particular, the uCTN model, post-compilation, is a 64-qubit circuit that contains 1207 gates, of which 360 are two-qubit entangling gates, which are two orders of magnitude noisier than single-qubit gates. Furthermore, H2-1's mid-circuit measurement and qubit-reset features allow for qubit reuse \cite{decross2022qubitreuse}. Remarkably, with qubit-reuse strategies, which essentially follow the logic of good tensor contraction strategies, the 64-qubit uCTN circuit is compressed down to an 11-qubit circuit.

\begin{figure}[t]
\centering
$$
\includegraphics[scale=1.0,align=c,draft=false]{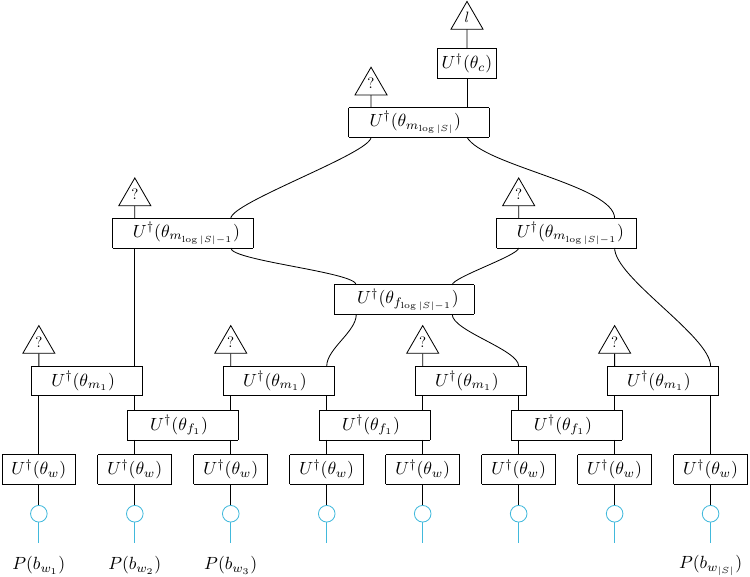}
$$
    \caption{A proposal for generative modelling for sequences. A non-syntactic circuit, like a CTN with $\bot={\tt discard}$ displayed here, could be trained to capture the joint distribution of sequences (leaves of the tree) and their labels (root of the tree), where the tokens are basis-encoded ($q=\log_2(|V|)$.
    Then, classification entails inputting a basis state in the leaves and sampling the root. Running the circuit in reverse, from root to leaves would implement a generator, where the basis state corresponding to the class $l$ would be input at the root of the tree, as well as random basis states denoted by "?" would be input at the branchings. One could actually prepare a mixed state (the operational reverse of the discard) by preparing a Bell state and discarding part of it.
    Then, the quantum state prepared on the leaves of the tree allows for sampling from a probability distribution over strings $P^l(b_{w_{1}}b_{w_{2}}\dots b_{w_{|S|}})$ that belong to the class $l$.
}
    \label{fig:gen}
\end{figure}

\section{Discussion and Outlook}


We have introduced machine learning models for sequence classification built in terms of tensor networks.
The architecture of the models plays the role of an inductive bias motivated by the inherent correlations and structures present in the data.
Moreover, the tensors are chosen to be unitary which allows for their instantiation as parameterised quantum circuits.
We have demonstrated both in simulation and by execution on a state-of-the-art trapped-ion quantum processor the efficient implementation of our models for datasets relevant to natural language processing as well as bioinformatics,
showing good performance.
The models and methods defined in this work enable for the first time the experimentation with a variety of quantum tensor network models for language processing. Importantly, with the methods presented here, one can employ large-scale real-world data, and explore the question of for what kinds of tasks and datasets is a syntax-aware structure beneficial, as well as explore more sophisticated model-selection strategies.
To this end, we make our code and results available in this repository\footnote{\url{https://github.com/CQCL/classification-with-qttn}}.


Further,
we aim to train \emph{quantum word embeddings}, or in general quantum token embeddings, and test their performance in a downstream task, such as the classification tasks tackled in this work.
This can be done with the usual objectives such as the skipgram \cite{skipgram} or glove \cite{glove} methods,
and it would be interesting to evaluate such quantum embeddings in tasks such as word analogy. Creating quantum word embeddings constitutes the creation of \emph{quantum data} to be classified; interestingly, it is in this context in which quantum convolutional neural networks (CTN with {\tt discard} in our work) were introduced \cite{cong}.
In addition, our methods can be straightforwardly extended so that a syntactic structure is learned using adaptive methods applicable to our STN or SCTN models \cite{lourens2023hierarchical},
which could be used to infer syntactic structures in bioinformatics data \cite{Searls2002}.
Our models can also be applied to other highly-structured data similarly displaying long-range correlations, such as neuron firing networks \cite{brainz}.

Beyond classification, our simple setup can be extended to accommodate generative modelling for sequences, using tree-like Born machines, using a setup as shown in Figure \ref{fig:gen}, by using a basis encoding of the vocabulary ($q=\log_2(|V|)$.
In general, one can create a quantum language model using one of our non-syntactic models, which is trained in a masking task to capture the conditional probability distribution of tokens in text corpora.
In addition, sampling the leaves of tree-like quantum circuits can be done polynomially faster on a quantum device.

Finally, we consider two general directions, guided by the efficiency of contracting tensor network models.
In this work, in our efficiently simulable models, we have only used the Born rule as a nonlinearity for obtaining the classification label, since we designed our models as to have valid quantum operational semantics.
In general, it is interesting to consider efficiently contractable TN models as `quantum-inspired' models to be executed on classical computers. Then one can use any element-wise nonlinearities one desires, acting on the complex-valued tensors \cite{ComplexValuedNNs2020}.
Alternatively, it would be interesting to consider text-level syntactic structures \cite{MathematicsTextStructure2021},
which are conjectured to lead to hard-to-contract tensor networks.
In that case, given quantum semantics, an efficient implementation would necessarily require quantum processors.

~\\~

\section*{Acknowledgements} CH acknowledges funding from grant EP/SO23607/1, and RY thanks Simon Harrison for his generous support for
the Wolfson Harrison UK Research Council Quantum Foundation Scholarship that he enjoys.
We thank Marcello Benedetti, Michael Lubasch, Michael Foss-Feig, Steve Clark, Dimitri Kartsaklis for discussions and feedback on the manuscript.
~\\~


\bibliographystyle{unsrt}
\bibliography{refs.bib}

\appendix 

~\\~
~\\~

{\bf \huge Appendix}


\section{Further model details}
\label{app:model-details}

\begin{figure}[H]
    \centering
$$
 \includegraphics[scale=1.0,align=c,draft=false]{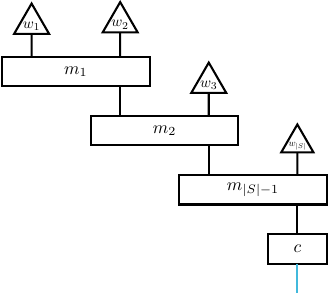} 
$$
    \caption{${\tt path}$.}
    \label{fig:path-app}
\end{figure}

\begin{figure}[H]
\centering
$$
 \includegraphics[scale=1.0,align=c,draft=false]{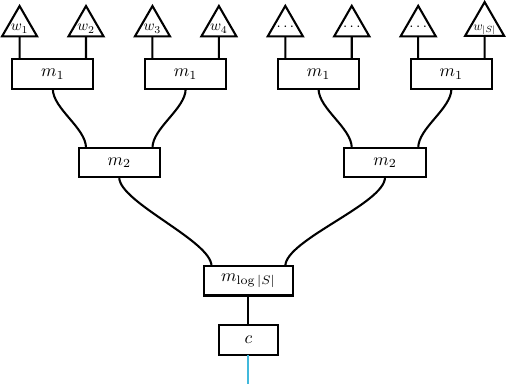} 
$$
    \caption{${\tt tree}$.}
    \label{fig:ttn-app}
\end{figure}

\begin{figure}[H]
\centering
$$
 \includegraphics[scale=1.0,align=c,draft=false]{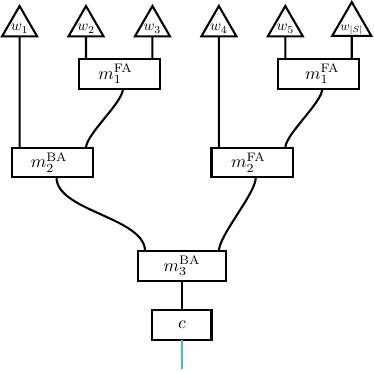} 
$$
    \caption{${\tt syntax}$, for a given CCG parse.}
    \label{fig:stn-app}
\end{figure}

\begin{figure}[H]
\centering
$$
\includegraphics[scale=1.0,align=c,draft=false]{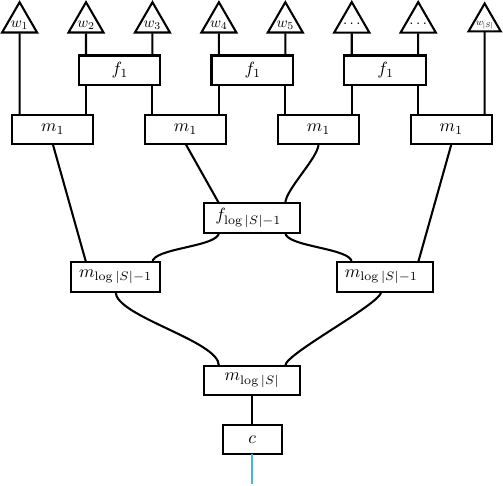} 
$$
    \caption{${\tt conv}$.}
    \label{fig:ctn-app}
\end{figure}

\begin{figure}[H]
    \centering
$$
\includegraphics[scale=1.0,align=c,draft=false]{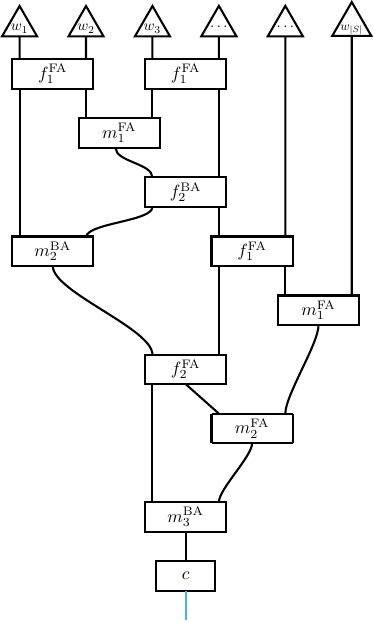} 
$$
    \caption{${\tt syntaxconv}$, for a given CCG parse.}
    \label{fig:sctn-app}
\end{figure}

The compositional {\tt schemes} used to instantiate the quantum tensor network models in Section \ref{sec:models} via the semantic functor $F$ are shown in Figures \ref{fig:path-app}, \ref{fig:ttn-app}, \ref{fig:stn-app}, \ref{fig:ctn-app}, and \ref{fig:sctn-app}.
These are all composed using the generators shown in Figure \ref{fig:functor} (left hand side of `$\mapsto$').
Black wires indicate $\tau$ types and blue wires indicate $\sigma$ types.
For every model species,
we show the complexity of full tensor contraction in Table \ref{tbl:complexities} and
the total number of parameters in
Table \ref{tbl:params-app}.

\begin{table}[H]
\centering
\small
\begin{tabular}{ |c||c| } 
\hline
Model species & $|\theta|$\\
\hline
uPTN & $ D  (  |\theta_w^l|  |V|
+   |\theta_m^l|
+   |\theta_c^l| )$\\
hPTN & $ D  (  |\theta_w^l|  |V|
+ |\theta_m^l| (|S|-1)
+ |\theta_c^l| )$\\
uTTN & $ D  (  |\theta_w^l|  |V|
+ |\theta_m^l|
+ |\theta_c^l| )$\\
hTTN & $ D  (  |\theta_w^l|  |V|
+ |\theta_m^l| \log(|S|)
+ |\theta_c^l| )$ \\
uSTN & $ D  (  |\theta_w^l|  |V|
+ |\theta_m^l|
+ |\theta_c^l| )$\\
rSTN & $\leq D (  |\theta_w^l|  |V|
+ |\theta_m^l| (|S|-1)
+ |\theta_c^l| )$\\
hSTN & $ \leq D  (  |\theta_w^l|  |V|
+ |\theta_m^l| |R|
+ |\theta_c^l| )$   \\
uCTN & $ D  (  |\theta_w^l| |V|
+ |\theta_m^l|  
+ |\theta_f^l| 
+ |\theta_c^l| )$\\
hCTN  & $ D  (  |\theta_w^l| |V|
+ |\theta_m^l| \log(|S|) 
+ |\theta_f^l| (\log(|S|)-1)
+ |\theta_c^l| )$ \\
uSCTN & $ D  (  |\theta_w^l| |V| + |\theta_m^l|  + |\theta_f^l|  + |\theta_c^l| )$\\
rSCTN & $ \leq D  (  |\theta_w^l| |V| + |\theta_m^l| |R|  + |\theta_f^l| |R| + |\theta_c^l| )$\\
hSCTN & $\leq D  (  |\theta_w^l| |V|+ |\theta_m^l| (|S|-1) + |\theta_f^l| (|S|-2) + |\theta_c^l| )$\\
\hline
\end{tabular}
\caption{Number of parameters per model species.}
\label{tbl:params-app}

\end{table}

\section{Methods}

\subsection{Efficient Parameterised Quantum Circuits}

The trainable tensors are given by circuit ansatz consisting of parameterized unitaries applied in parallel and sequence to qubits. Substitution of the parameters to the resulting symbolic functions can be very slow and inhibiting for large data training. However, we can get around this using JAX. Instead of constructing a symbolic function and substituting our updated parameters after each training iteration we can instead JAX compile a pure function that takes the parameter values as input and applies the numerical gate unitaries directly. This construction uses DisCoPy's quantum tensor network library. We then batch this operation using \texttt{jax.vmap} for further efficiency.

  
\subsection{Batched contraction of PTN, TTN, STN}

In order to run this model efficiently for large-scale data it is necessary to implement batching of the trees which, is non-trivial for the varying structures present in the syntactic model. We achieve this by padding the sentences to be of equal length, $N$, such that all trees have $N-1$ merge operations. We then define a length $N$ vector for each tree where each element of this vector is either a classical vector, an initialised word state or a null pad state. When $\bot=\texttt{discard}$, all quantum states are set to \texttt{mixed} such that density matrices and Kraus operators are used throughout. The syntax trees in our batch provide an ordering, a list of pairs of indices corresponding to the $N$-vectors, in which we need to apply the \texttt{merge}  
operations. We define a \texttt{merge} function that takes as input the relevant parameters defining this operation and the two elements taken from the $N$-vector according to the current tree indices. After the pooling-like operation we update the $N$-vector at the lower index with this result ready for it to be merged again later in the tree or measured. The pad indices are simply ($N-1$, $N$) as the final classification state will always be at index 0 and hence updating the pad indices will not affect results. So the $i$-th \texttt{merge} operation can now be applied in parallel across all trees, again using \texttt{jax.vmap}, and we sweep over these operations sequentially until all $N-1$ are performed utilising \texttt{jax.lax.scan}. This allows for fast batched contractions which are easily run on GPUs for further efficiency. Discarding and measurement are all readily implemented using \texttt{DisCoPy}.

\subsection{Batched contraction of CTN, SCTN}

Due to the presence of disentanglers, the CTN calculation cannot be factorised as above so instead JIT compilation and \texttt{jax.vmap} is applied to the full contraction sequence. Additionally, the CTN must be padded to a length of $2^n$, for integer $n$, so we pad to the minimum above the true length of the sequence. Batched contraction is then applied to this grouped data. For the SCTN, inability to factorise the contraction is significantly more limiting for simulation of these networks. Saving a JIT compiled model for all syntactic structures is too memory extensive. Thus, to present results for this model we select the 100 most frequently occurring syntactic structures present in the data and batch over the examples with shared structure.

\end{document}

%% file: tree.tikzstyles
\tikzstyle{new style 0}=[fill={rgb,255: red,255; green,43; blue,199}, draw=black, shape=circle]
\tikzstyle{new style 1}=[fill={rgb,255: red,237; green,255; blue,38}, draw=black, shape=circle]
\tikzstyle{new style 2}=[fill=white, draw={rgb,255: red,65; green,185; blue,220}, shape=circle]
\tikzstyle{state}=[shape=isosceles triangle, isosceles triangle apex angle=60, draw=black, inner sep=0.8pt, minimum width=0.75cm, minimum height=5mm, shape border rotate=90, text depth=1mm, tikzit fill={rgb,255: red,154; green,1; blue,3}, scale=0.85]
\tikzstyle{effect}=[shape=isosceles triangle, isosceles triangle apex angle=60, draw=black, inner sep=0.8pt, minimum width=0.75cm, minimum height=5mm, shape border rotate=-90, text depth=1mm, tikzit fill={rgb,255: red,12; green,154; blue,41}, scale=0.85]

\tikzstyle{new edge style 0}=[-, draw={rgb,255: red,65; green,185; blue,220}]
\tikzstyle{new edge style 1}=[-]
\tikzstyle{thick}=[-, line width=1]